\documentclass[conference]{IEEEtran}
\IEEEoverridecommandlockouts
\usepackage{cite}
\usepackage{amsmath,amssymb,amsfonts}
\usepackage{algorithmic}
\usepackage{graphicx}
\usepackage{textcomp}
\usepackage{xcolor}
\usepackage{booktabs} 
\usepackage{placeins}     
\usepackage{dblfloatfix}  

\def\BibTeX{{\rm B\kern-.05em{\sc i\kern-.025em b}\kern-.08em
    T\kern-.1667em\lower.7ex\hbox{E}\kern-.125emX}}
    
\begin{document}

\title{SSEMG-NET: A Spectrogram-Based Mamba Network for Surface Electromyography Denoising}

\author{
    \IEEEauthorblockN{
        Cheng-Han Shih\textsuperscript{1,3*}, 
        Kuan-Chen Wang\textsuperscript{2,3*}, 
        Kai-Chun Liu\textsuperscript{4,5}, 
        Ping-Cheng Yeh\textsuperscript{2}, and 
        Yu Tsao\textsuperscript{3}
    }
    \IEEEauthorblockA{
        \textsuperscript{1}Department of Biomedical Engineering, National Taiwan University, Taiwan \\
        \textsuperscript{2}Graduate Institute of Communication Engineering, National Taiwan University, Taiwan \\
        \textsuperscript{3}Research Center for Information Technology Innovation, Academia Sinica, Taiwan \\
        \textsuperscript{4}Department of Mechanical Engineering, Stevens Institute of Technology, USA \\
        \textsuperscript{5}Dept. of Rehab. and Regenerative Medicine, Columbia University Irving Medical Center, USA
    }
    \IEEEauthorblockA{
        Email: b12508006@ntu.edu.tw, d12942016@ntu.edu.tw, kliu33@stevens.edu, pcyeh@ntu.edu.tw, yu.tsao@citi.sinica.edu.tw
    }
    \thanks{*These authors contributed equally to this work.}
}

\maketitle

\begin{abstract}
Electrocardiogram (ECG) artifact contamination frequently occurs in surface electromyography (sEMG) when muscles are recorded near the heart. Existing neural network (NN)-based approaches typically perform waveform-level end-to-end denoising with pointwise losses, but often fail to preserve the spectral structures of sEMG. In addition, these models do not exploit the prior knowledge that ECG mainly distorts the low-frequency band of sEMG. We propose SSEMG-Net, a spectrogram-domain model that explicitly captures sEMG–ECG interactions using a bidirectional Mamba backbone and a dual-head decoder for magnitude masking and wrapped-phase estimation. The magnitude masking module is physiologically aware, restricting denoising to the low-frequency sub-band where ECG activity predominates. Experimental results demonstrate that SSEMG-Net achieves superior signal quality and lower feature-extraction error compared to prior methods, underscoring its potential for clinical applications where preserving sEMG spectral features is essential.
\end{abstract}


\section{Introduction}
\label{sec:intro}
Surface electromyography (sEMG) provides a non-invasive measure of neuromuscular activity by recording the electrical potentials generated during muscle contractions. It has been widely applied in neuromuscular research~\cite{tang2018novel}, rehabilitation~\cite{engdahl2015surveying}, stress and fatigue monitoring~\cite{wijsman2013wearable}, clinical evaluation of neuromuscular and respiratory disorders~\cite{domnik2020clinical}, prosthesis control~\cite{ma2014hand}, and gesture recognition in virtual reality~\cite{cote2021transferable}. However, when the measured muscles are in the proximity of the heart, sEMG signals can be contaminated by electrocardiogram (ECG) activity~\cite{merletti2020tutorial}. Such contamination distorts both amplitude and spectral features, which in turn compromises physiological analyses and degrades the performance of downstream applications. Hence, effective ECG artifact removal is essential.

Removing ECG artifacts is challenging because the main frequency bands of ECG (about 0–150 Hz) and sEMG (about 10–500 Hz) overlap~\cite{winter2009biomechanics}. Conventional single-channel approaches, such as high-pass (HP) filters and template subtraction (TS), can reduce ECG interference but suffer from clear drawbacks.  HP filters remove low-frequency ECG but also eliminate important sEMG components, while TS assumes quasi-periodic ECG patterns and zero-mean Gaussian sEMG, which are rarely satisfied in practice~\cite{merletti2020tutorial}. 

Neural networks (NNs) have recently shown promise in sEMG denoising due to their nonlinear modeling capabilities. For example, fully convolutional networks (FCNs) outperform HP and TS approaches across a broad range of SNRs ~\cite{wang2023ecg}. SDEMG, a score-based diffusion model, further improves signal fidelity by modeling underlying data distributions, but with high computational cost~\cite{liu2024sdemg}. More recently, MSEMG combined convolutional encoders with the Mamba state-space model, effectively capturing long-range dependencies while maintaining efficiency~\cite{liu2025msemg}.

Despite recent advances, NN-based sEMG denoising methods still face challenges preserving spectral fidelity. A key limitation stems from the end-to-end waveform mapping paradigm: with limited training data and complex waveform structures, models can have difficulty suppressing low-frequency ECG artifacts while retaining high-frequency sEMG information. Moreover, most approaches rely solely on pointwise loss functions (e.g., L1 or L2), which enforce sample-level accuracy but do not explicitly constrain spectral structure~\cite{germain2019speechdenoising, wang2025trustemg}. These shortcomings are problematic for applications that depend on frequency fidelity, such as muscle fatigue assessment~\cite{cifrek2009surface}. This underscores the need for an sEMG denoising framework that jointly preserves both temporal and spectral fidelity.

In this study, we propose SSEMG-Net, a spectrogram-based framework for ECG artifact removal in sEMG. The model operates in the time–frequency (TF) domain, explicitly reconstructing both magnitude and phase. For magnitude, it employs a physiologically aware sub-band mask that targets 0–150 Hz, where ECG contamination is prominent, while preserving mid- to high-frequency sEMG content. To further enhance reconstruction, we adopt composite time- and frequency-domain objectives, including multi-resolution short-time Fourier transform (STFT) loss. Experimental results show that SSEMG-Net consistently outperforms prior methods, achieving robust denoising across SNR levels and superior preservation of spectral features. To our knowledge, this is the first NN-based sEMG enhancement framework to explicitly perform denoising in the TF domain.

\begin{figure*}[tbh!]
  \centering
  \includegraphics[width=.9\textwidth]{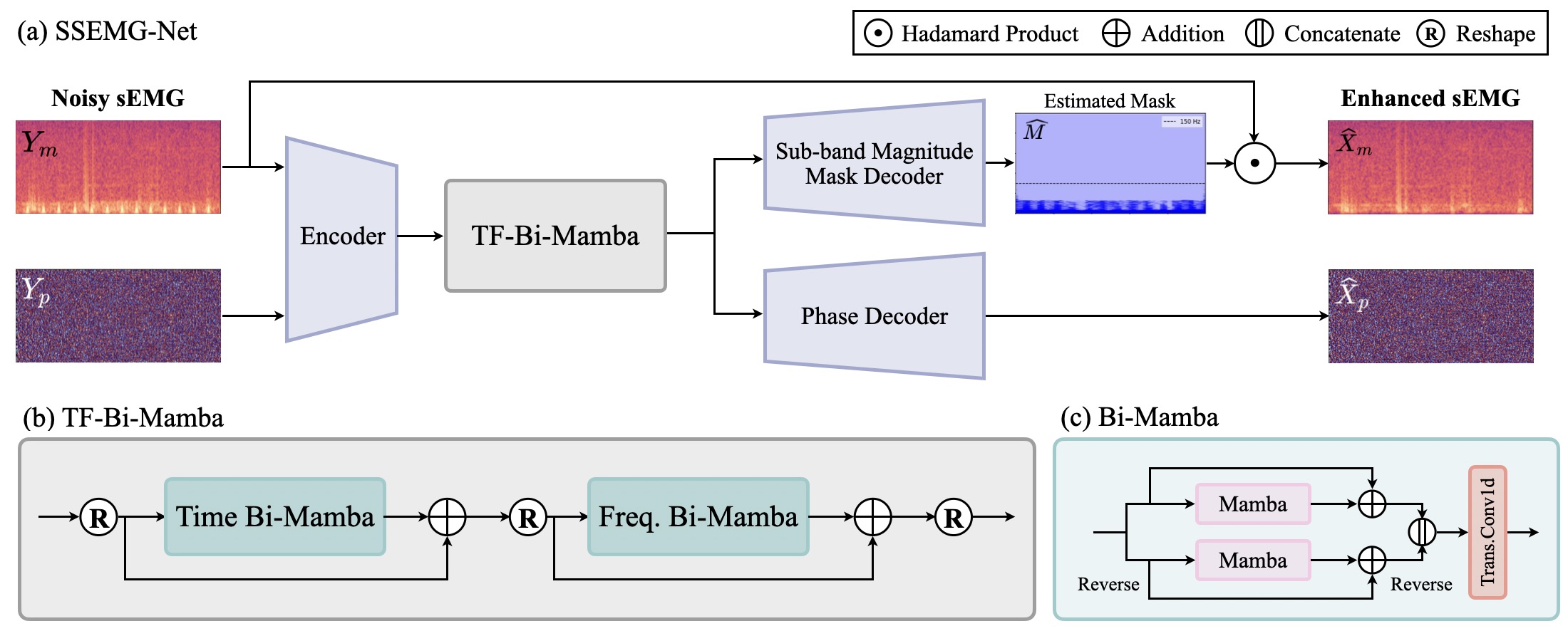}
  \caption{The architectures of (a) the proposed SSEMG-Net, (b) TF-Bi-Mamba module, and (c) Bi-Mamba block.}
  \label{fig:arch}
\end{figure*}

\section{Related Work}
\label{sec:related}
\subsection{ECG interference removal methods}
\label{ssec:previous}
Conventional ECG removal methods, such as HP filters and TS, eliminate ECG artifacts from sEMG signals using spectral- and temporal-domain processing, respectively. HP filters suppress ECG interference by discarding the low-frequency portion of the signal, but they inevitably remove low-frequency sEMG components as well. TS, on the other hand, identifies ECG waveforms and subtracts averaged or filtered ECG templates. While effective under controlled conditions, TS relies on assumptions of quasi-periodic ECG morphology and zero-mean Gaussian sEMG, which often do not hold in real-world recordings.

Recently, NN–based methods have emerged as powerful alternatives. Wang et al.~\cite{wang2023ecg} introduced an FCN denoising autoencoder for ECG artifact removal from sEMG waveforms, which consistently outperformed HP and TS across different noise levels. Liu et al.~\cite{liu2024sdemg} developed SDEMG, a score-based diffusion model that enhanced sEMG signal fidelity by capturing the underlying distribution of clean sEMG at considerable computational cost. Liu et al.~\cite{liu2025msemg} further introduced MSEMG by incorporating the Mamba state-space model into a convolutional architecture to better capture long-range dependencies, achieving higher denoising performance with fewer parameters. Despite their differences, these NN-based approaches share a common design: waveform-based end-to-end denoising with a pointwise objective function. While effective in suppressing ECG artifacts, this framework is often limited in preserving fine-grained sEMG characteristics, particularly in the spectral domain.

\subsection{Spectrogram-based denoising neural networks}
\label{ssec:mamba}
Spectrogram-based denoising NNs have recently been explored for bio-signal denoising \cite{hung2024mecg,pei2024dtp,lim2025deep}. By operating in the TF-domain, these approaches can leverage both temporal dynamics and spectral patterns, providing more precise noise suppression and better preservation of frequency content compared to end-to-end waveform methods. A representative work in this direction is MECG-E~\cite{hung2024mecg}, developed for ECG denoising and baseline wander removal. MECG-E adopts the Mamba state-space model as its backbone and processes STFT representations of ECG signals. To improve robustness and ensure spectral consistency, it incorporates a composite training objective that combines time-domain loss with spectral-domain losses. MECG-E achieves state-of-the-art performance in ECG denoising and demonstrates efficient inference compared to diffusion-based approaches, highlighting the potential of spectrogram-domain modeling for bio-signals. This motivates our proposed SSEMG-Net, which extends the spectrogram-based denoising framework to ECG removal from sEMG.

\section{Methodology}
\label{sec:proposed}

\label{ssec:frontend}
Fig. \ref{fig:arch} shows the structure of SSEMG-Net. A noisy sEMG segment \(y\in\mathbb{R}^{1\times L}\) is first converted by STFT into a complex time-frequency (TF) representation \(Y\in\mathbb{R}^{T\times F\times 2}\), where \(L\),  \(T\), and \(F\) denote the length of the sEMG signal, the number of time frames, and the number of frequency bins, respectively. To stabilize optimization and compress dynamic range, we apply a power-law to the magnitude while preserving the phase:
\begin{equation}
    Y^c=\lvert Y\rvert^{\,c} e^{jY_p}
    = Y_m^c e^{jY_p},
    \label{eq:powerlaw}
\end{equation}
where \(c\) is the magnitude compression exponent, and \(Y_m^c, Y_p\in\mathbb{R}^{T\times F}\) denote the compressed magnitude and wrapped phase, respectively.

\subsection{Model architecture}
\label{ssec:arch}

\subsubsection{Encoder}
The encoder comprises two convolutional blocks separated by a dilated DenseNet~\cite{mao2025dilated}. The first block expands channel capacity with group normalization~\cite{wu2018groupnormalization} and PReLU activation~\cite{godfrey2019parametric}, while the DenseNet enlarges the temporal receptive field and preserves temporal--spectral details. The second block downsamples along frequency to reduce computational cost, yielding a TF-domain representation with extended temporal context and compact frequency resolution.  

\subsubsection{Time--Frequency bidirectional mamba}
As shown in Fig.~\ref{fig:arch} (b), the TF-Bi-Mamba consists of a Time bidirectional Mamba (Bi-Mamba) that models bidirectional dynamics along the temporal axis and a Frequency Bi-Mamba that captures cross-frequency relations, jointly modeling temporal and spectral structure~\cite{hung2024mecg}. Each block processes sequences bidirectionally, with forward and reversed inputs fused through residual and transposed convolution layers to capture past--future context, as shown in Fig.~\ref{fig:arch} (c). Mamba layers are implemented as state-space models augmented with depthwise convolutions and RMS normalization, enabling efficient modeling of long-range dynamics and localized spectral structure. This configuration enables the model to jointly capture long-range temporal dynamics and localized spectral structure, yielding TF-domain representations well suited for magnitude masking and phase estimation.

\subsubsection{Decoder}
SSEMG-Net employs two decoders, a Sub-band Magnitude Mask Decoder and a Phase Decoder, inspired by~\cite{lu2023mpsenet}. The Sub-band Magnitude Mask Decoder consists of a dilated DenseBlock, an upsampling transposed convolution, and a \(1{\times}1\) projection with InstanceNorm and PReLU. It generates a mask $\widehat{M} \in \mathbb{R}^{T\times F}$ with the same dimension as the magnitude \(Y_{m}^{c}\), from which the enhanced magnitude $\widehat{X}_m^{c}$ is obtained as
\begin{equation}
\widehat{X}_m^{c} = Y_{m}^{c} \odot \widehat{M},
\label{eq:enhanced_mag}
\end{equation}
where \(\odot\) denotes the Hadamard product. To preserve high-frequency sEMG components, the mask $\widehat{M}$ is fixed to one above a cutoff frequency $f_{\text{cut}}$, and learned parametrically below $f_{\text{cut}}$ to conduct denoising in the sub-band where ECG overlaps and dominates with sEMG.
\begin{equation}
\widehat{M}(t,f) =
\begin{cases}
\sigma\!\big(Z(t,f)\big), & f < f_{\text{cut}}, \\[4pt]
1, & f \ge f_{\text{cut}},
\end{cases}
\label{eq:gmask_piecewise}
\end{equation}
where $Z(t,f)$ is the scalar pre-activation at TF coordinate $(t,f)$, and $\sigma(\cdot)$ is a learnable sigmoid function~\cite{fu2021metricganplus}.

The Phase Decoder consists of a dilated DenseNet, a deconvolutional block, and a parallel phase estimation architecture. The dilated DenseNet refines the TF features from the TF-Bi-Mamba, enlarging the receptive field while preserving detailed temporal and spectral information. This is followed by a transposed convolutional layer that restores the original frequency resolution. The parallel architecture then employs two \(1\times1\) convolutional layers to generate the pseudo-real \(\widehat{R}\) and pseudo-imaginary \(\widehat{I}\) components, which represent the unnormalized Cartesian coordinates of a unit phasor encoding the phase direction. The wrapped phase spectrum is subsequently estimated using the two-argument arctangent function:
\begin{equation}
\widehat{X}_p=\operatorname{atan2}\!\big(\widehat{I},\,\widehat{R}\big).
\label{eq:phase_atan2}
\end{equation}

Combining the enhanced magnitude \(\widehat{X}_m\) (restored by the inverse power-law) with the wrapped phase \(\widehat{X}_p\), the final enhanced complex spectrogram can be expressed as:
\begin{equation}
\widehat{X} = \big(\widehat{X}_m^{c}\big)^{1/c}\,e^{j\widehat{X}_p} = \widehat{X}_m\,e^{j\widehat{X}_p}.
\label{eq:synthspec}
\end{equation}

\subsection{Loss function}
\label{ssec:loss}
The loss function consists of four terms: a time-domain loss $\mathcal{L}_{{time}}$, a complex-domain loss $\mathcal{L}_{{cpx}}$, a consistency loss $\mathcal{L}_{{con}}$, and a multi-resolution STFT (MR-STFT) loss $\mathcal{L}_{{mr}}$. The first three terms follow MECG-E~\cite{hung2024mecg}:$\mathcal{L}_{{time}}$ enforces waveform fidelity, $\mathcal{L}_{{cpx}}$ constrains reconstruction in the complex spectrogram domain, and $\mathcal{L}_{{con}}$ mitigates STFT and inverse STFT inconsistencies. These terms are calculated as:
\begin{align}
  \mathcal{L}_{{time}} &=
    \lVert x-\hat{x}\rVert_{1}, \label{eq:ltime}\\
  \mathcal{L}_{{cpx}} &=
    \lVert X-\hat{X}\rVert_{2}^{2}, \label{eq:lcpx}\\
  \mathcal{L}_{{con}} &=
    \big\lVert \hat{X}-\mathrm{S}_{\boldsymbol{\theta}_{0}}(\hat{x}) \big\rVert_{2}^{2}, \label{eq:lcon}
\end{align}
where $x$, $\hat x$, $X$, and $\hat X$ denote the clean waveform, enhanced waveform, clean complex spectra, and enhanced complex spectra, respectively. $\mathrm{S}_{\boldsymbol{\theta}_{0}}(\cdot)$ denotes the STFT with analysis parameters $\boldsymbol{\theta_{0}}$, including $(N_{\text{fft}},\text{Hop},\text{Window size})$. 

Building on this, we introduce $\mathcal{L}_{{mr}}$ to provide supervision across multiple analysis resolutions of the spectrogram ~\cite{yamamoto2020parallel}.

\begin{align}
  \mathcal{L}_{{mr}}
  = \frac{1}{R} \sum_{r=1}^{R} \Big(
    &\ \big\lVert\,|\mathrm{S}_{\boldsymbol{\theta}_r}(x)|
      - |\mathrm{S}_{\boldsymbol{\theta}_r}(\hat{x})|\,\big\rVert_{1} \notag \\
    &+ \big\lVert \mathrm{S}_{\boldsymbol{\theta}_r}(x)
      - \mathrm{S}_{\boldsymbol{\theta}_r}(\hat{x}) \big\rVert_{2}^{2}
  \Big),
  \label{eq:mrloss}
\end{align}
where $R$ is the number of STFTs with different analysis parameters $\theta_r$. By using MR-STFT losses, the model can learn the time-frequency characteristics of sEMG and avoid overfitting to a fixed STFT representation, which can lead to suboptimal results in waveform reconstruction.
The total loss function is expressed as:
\begin{equation}
  \mathcal{L}_{{all}}
  = \gamma_1 \mathcal{L}_{{time}}
  + \gamma_2 \mathcal{L}_{{cpx}}
  + \gamma_3 \mathcal{L}_{{con}}
  + \gamma_4 \mathcal{L}_{{mr}},
  \label{eq:allloss}
\end{equation}
where $\gamma_1$, $\gamma_2$, $\gamma_3$, and $\gamma_4$ are the weights for the losses.

\section{Experiments}
\subsection{Dataset preprocessing and preparation}
This study utilized sEMG data from the DB2 subset of Non-Invasive Adaptive Prosthetics (NINAPro) database \cite{atzori2014electromyography}, comprising 12-channel sEMG recordings from 40 subjects performing hand movement across three sessions (Exercises 1--3, with 17, 22, and 10 movement classes, respectively). Each movement was repeated six times consecutively for 5 s and then followed by a 3-s rest. This database has served as a clean sEMG source after proper filtering in previous studies \cite{MACHADO2021102752}. sEMG signals were preprocessed with a fourth-order Butterworth bandpass filter (20--500 Hz), downsampled to 1 kHz, normalized, and segmented into 10-s segments.

For ECG interference, we adopted the MIT-BIH Normal Sinus Rhythm Database (NSRD) \cite{goldberger2000physiobank}, providing 2-channel ECG recordings collected from 18 healthy subjects at a sampling rate of 128 Hz. The Channel 1 recordings were preprocessed using third-order Butterworth high-pass (10 Hz) and low-pass filters (200 Hz) to suppress potential noise. This database has been employed in prior work for modeling ECG contamination in sEMG \cite{wang2023ecg,liu2024sdemg,liu2025msemg}. 
 
We used sEMG data from Channel 2, Exercises 1 and 3, 30 subjects, for training and validation. For each training sample, 10 ECG signals were randomly chosen from 12 NSRD subjects and superimposed onto the sEMG as interference at six SNR levels ranging from –15 to –5 dB in 2 dB steps. The validation set used ECG data from three other subjects with the same SNRs. To assess generalization, testing was performed on different subjects, movements, channels, and SNRs. Specifically, sEMG data were from Channels 9–12 of Exercise 2, 10 subjects, while ECG signals from three remaining NSRD subjects (record IDs 16420, 16539, and 16786) were superimposed at SNR levels from –14 to 0 dB in 2 dB steps.

\subsection{Implementation details}
SSEMG-Net was trained for 30 epochs with AdamW (learning rate \(3\times10^{-4}\)), mini-batch size 4, and gradient accumulation of 4 (effective batch size 16). The STFT front end used a Hann window of 512 samples and a hop of 128 at 1 kHz; the magnitude compression exponent was \(c{=}0.5\). The encoder employed 64 channels with GroupNorm (8 groups) and PReLU; its dilated DenseNet had depth 4 with dilations \(\{1,2,4,8\}\). The mask decoder applied a learnable per-frequency sigmoid with scaling $\alpha_f$. Each $\alpha_f$ was parameterized via a trainable raw variable with a softplus transformation and a $0.05$ offset to ensure strictly positive slopes and stable training. We set $\beta=1.0$ and enforced a deterministic pass-through above the cutoff frequency $f_{\text{cut}}=150\,\text{Hz}$. The weights $\gamma_1$, $\gamma_2$, $\gamma_3$, and $\gamma_4$, in eq. \ref{eq:allloss} are set to \((0.25,\,0.40,\,0.25,\,0.08)\); \(\mathcal{L}_{mr}\) was computed at three sets of parameters: \((256,64,256)\), \((512,128,512)\), and \((1024,256,1024)\).We set the masking cutoff frequency to $f_{\text{cut}}=150\,\text{Hz}$ by default.

\subsection{Evaluation metrics}
Performance was evaluated based on signal reconstruction
quality and feature extraction accuracy~\cite{xu2020comparative,drake2006elimination,liu2024sdemg,liu2025msemg}.  
For signal reconstruction quality, we report SNR improvement (SNR$_{{imp}}$) and root-mean-square error (RMSE), which quantify the gain in SNR relative to the contaminated input and the deviation of the denoised waveform from the clean reference, respectively.  
For feature extraction error, we compute the RMSE of the average rectified value (ARV) and the mean frequency (MF) feature vectors. Notably, MF denotes the centroid of the spectrum, which can represent spectral fidelity. 
All metrics follow established protocols in prior work~\cite{xu2020comparative,wang2023ecg,liu2024sdemg}. Higher SNR$_{{imp}}$ values, together with lower RMSE, RMSE$_{{ARV}}$, and RMSE$_{{MF}}$, indicate more effective ECG artifact suppression and better preservation of sEMG characteristics.

\section{Results and Discussion}

\subsection{Overall comparison with prior methods}
Table~\ref{tab:overall_performance} compares SSEMG-Net with representative single-channel baselines and recent neural network methods for ECG artifact removal in sEMG. 
The baselines include classical single-channel preprocessing methods such as high-pass filtering and template subtraction. 
These methods can suppress ECG artifacts but they can also remove low-frequency neuromuscular information~\cite{merletti2020tutorial}. 
The baselines also include data-driven waveform models that learn nonlinear denoising under ECG and sEMG spectral overlap~\cite{wang2023ecg,liu2024sdemg,liu2025msemg}. 
SSEMG-Net achieves the strongest performance across all reported metrics. 
This result indicates that TF-domain modeling improves reconstruction quality and feature integrity under ECG contamination.

A key outcome is the consistent reduction in feature-level errors, especially RMSE$_{MF}$. 
Mean frequency is a common frequency-domain biomarker in neuromuscular assessment. 
It is widely used for fatigue-related analysis~\cite{cifrek2009surface}. 
Small spectral biases can propagate into downstream estimation errors. 
They can also change the physiological interpretation~\cite{cifrek2009surface,merletti2020tutorial}. 
The improvement in RMSE$_{MF}$ supports the claim that SSEMG-Net preserves the spectral distribution of sEMG. 
This preservation is important because downstream feature extractors depend on spectral fidelity.

\begin{figure}[!t]
  \centering
  \includegraphics[width=\linewidth]{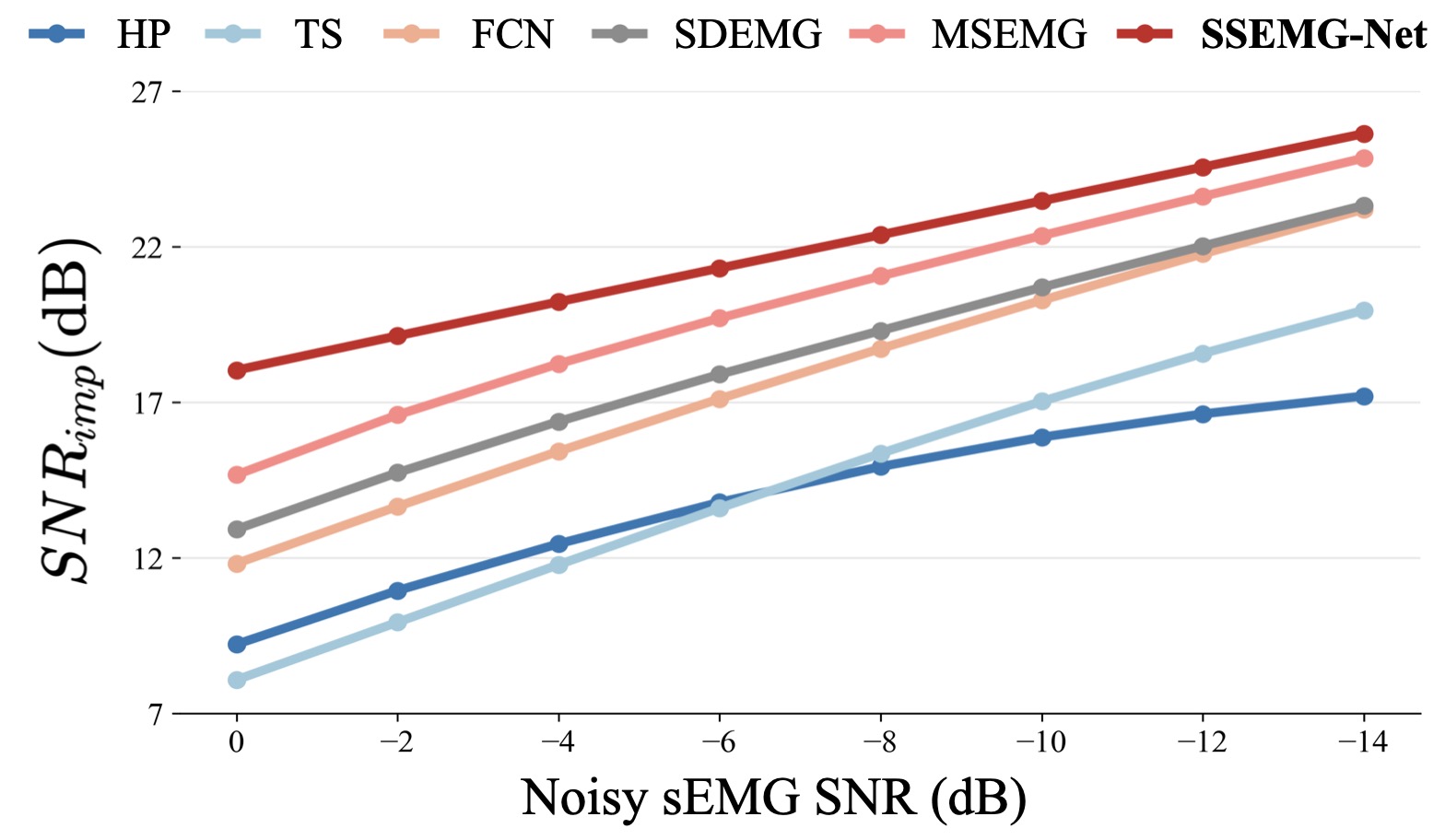}
  \caption{Comparison of SNR$_{imp}$ performance under different SNR inputs.}
  \label{fig:snr_curve}
\end{figure}

\begin{figure}[!t]
  \centering
  \includegraphics[width=\linewidth]{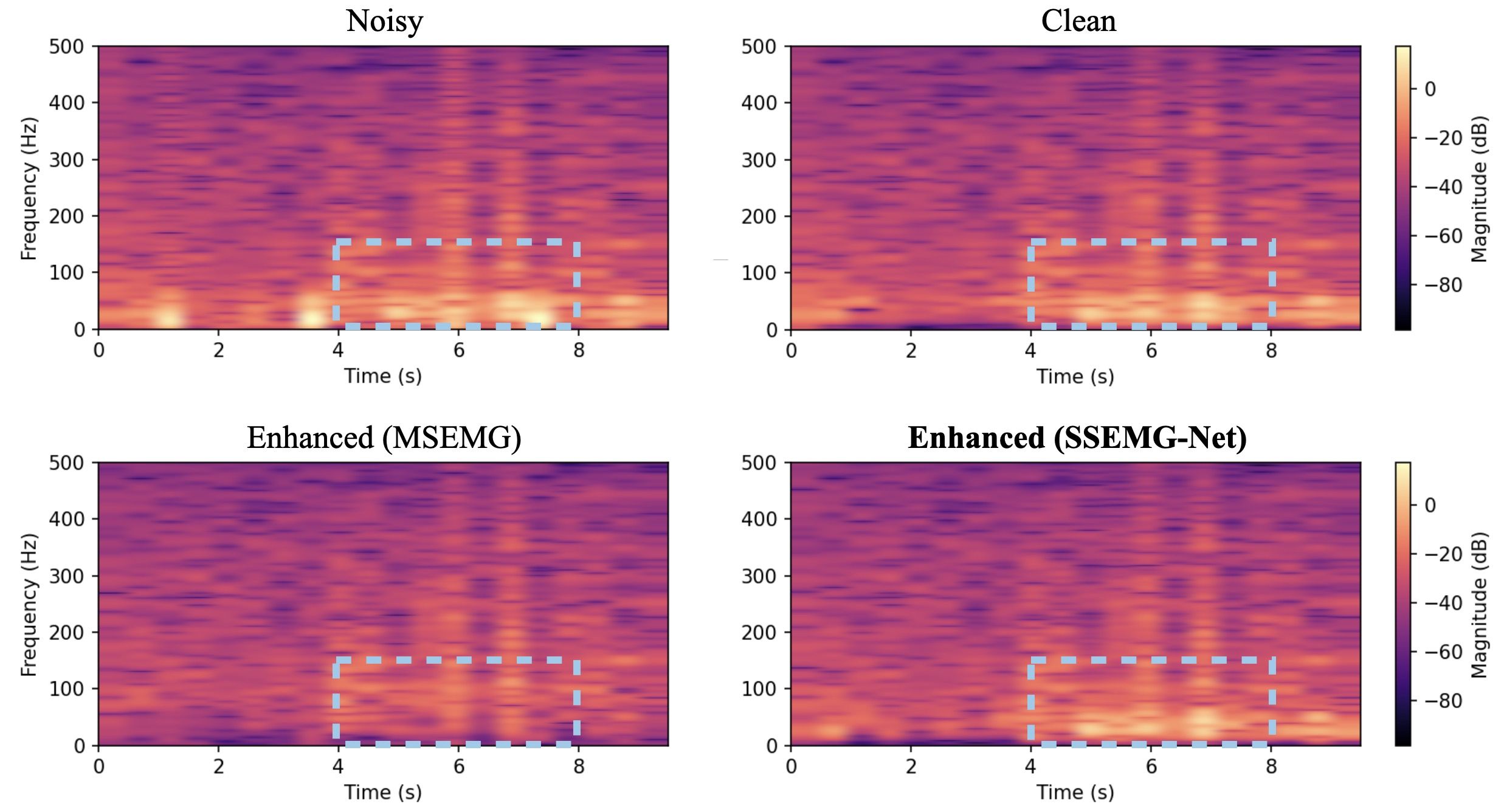}
  \caption{Comparison of enhanced sEMG spectrograms by SSEMG-Net and MSEMG. The blue rectangle marks the region where SSEMG-Net more closely reconstructs the clean reference compared to MSEMG.}
  \label{fig:qual_spec}
\end{figure}

Fig.~\ref{fig:snr_curve} shows that SSEMG-Net outperforms competing approaches across a wide range of input SNRs. 
This pattern indicates robust denoising under both severe and mild contamination. 
Fig.~\ref{fig:qual_spec} provides qualitative evidence from spectrogram comparisons. 
SSEMG-Net reconstructs TF structures more faithfully than a strong waveform baseline. 
It reduces residual low-frequency interference. 
It does not introduce visible broadband spectral warping.

\begin{table}[!t]
\centering
\small
\renewcommand\arraystretch{1.25}
\caption{Overall performance comparison of previous methods and SSEMG-Net.}
\label{tab:overall_performance}
\resizebox{\linewidth}{!}{%
\begin{tabular}{lcccc}
\toprule
 & SNR\(_{imp}\) (dB) ↑ & RMSE ↓ & RMSE\(_{ARV}\) ↓ & RMSE\(_{MF}\) (Hz) ↓ \\
\midrule
HP \cite{xu2020comparative}            & 13.885 & 1.735e$-$2 & 3.064e$-$3 & 19.471 \\
TS \cite{drake2006elimination}        & 14.279 & 1.626e$-$2 & 3.859e$-$3 & 23.149 \\
FCN \cite{wang2023ecg}                & 17.758 & 1.178e$-$2 & 3.864e$-$3 & 18.038 \\
SDEMG \cite{liu2024sdemg}             & 18.467 & 1.138e$-$2 & 2.809e$-$3 & 14.435 \\
MSEMG \cite{liu2025msemg}             & 20.317 & 8.603e$-$3 & 2.382e$-$3 & 11.379 \\
\textbf{SSEMG-Net (proposed)}         & \textbf{22.138} & \textbf{7.565e$-$3} & \textbf{1.004e$-$3} & \textbf{8.003} \\
\bottomrule
\multicolumn{5}{l}{*Bold font indicates the best value for each metric.}
\end{tabular}}
\end{table}

\subsection{Ablation studies}
\label{ssec:ablation}
The ablation results are intended to justify the main design choices in SSEMG-Net. 
All experiments use the same data construction, training schedule, and evaluation pipeline. 
Each row in Table~\ref{tab:ablation_main} changes one element of the system while keeping the rest fixed. 
This setup makes the observed differences attributable to the targeted factor rather than to confounded training effects.

Table~\ref{tab:ablation_main} shows that both TF reconstruction constraints and the physiologically guided sub-band design contribute to feature-faithful denoising. 
Removing MR-STFT loss reduces performance across metrics. 
This indicates that pointwise or single-resolution supervision alone does not sufficiently constrain spectral structure. 
Multi-resolution STFT supervision encourages agreement across analysis scales and reduces resolution-specific fitting that can bias frequency-domain features~\cite{yamamoto2020parallel}. 
In this task, spectral bias is reflected by MF-related error because MF shifts when residual low-frequency energy remains or when the spectrum is reshaped.

Removing explicit phase estimation causes the largest degradation. 
This result is important because ECG and sEMG overlap in the TF domain. 
Magnitude suppression alone can still yield inconsistent reconstruction after inverse STFT. 
Phase errors can distort the waveform and redistribute spectral energy, which leads to MF drift and larger feature error. 
This matters in biomedical applications because MF and related spectral descriptors are used to quantify neuromuscular state, including fatigue-related changes~\cite{cifrek2009surface}. 
The ablation supports phase estimation as a necessary component for feature-faithful denoising.

The comparison between the proposed sub-band mask and the w/o sub-band variant further supports the physiological inductive bias. 
The sub-band mask restricts learnable suppression to the low-frequency region where ECG interference dominates and keeps the remaining band unchanged. 
This design directly matches the clinical objective of ECG artifact removal in sEMG. 
It protects mid and high frequencies that carry neuromuscular information. 
It also limits avoidable spectral reshaping that can compromise interpretability~\cite{merletti2020tutorial}. 
The w/o sub-band variant allows the model to modify the entire spectrum. 
This added freedom can slightly improve pointwise reconstruction metrics. 
However, it can also attenuate or reshape sEMG-dominant frequencies that should be preserved. 
This mechanism explains why waveform-level gains do not necessarily translate into better MF fidelity. 
For this task, preserving biomarker validity is more important than optimizing pointwise error.

\begin{table}[!t]
\centering
\small
\renewcommand\arraystretch{1.25}
\caption{Ablation analysis of loss functions, phase estimation, and masking strategy.}
\label{tab:ablation_main}
\resizebox{\linewidth}{!}{%
\begin{tabular}{lcccc}
\toprule
 & SNR\(_{imp}\) (dB) ↑ & RMSE ↓ & RMSE\(_{ARV}\) ↓ & RMSE\(_{MF}\) (Hz) ↓ \\
\midrule
\textbf{SSEMG-Net (proposed)}      
& 22.138
& 7.565e$-$3
& \textbf{1.004e$-$3}
& \textbf{8.003} \\
\textit{-- w/o} MR-STFT loss       
& 21.164 & 8.295e$-$3 & 1.794e$-$3 & 14.303 \\
\textit{-- w/o} phase estimation   
& 19.696 & 9.878e$-$3 & 1.422e$-$3 & 17.673 \\
\textit{-- w/o} sub-band mask      
& \textbf{22.203} & \textbf{7.472e$-$3} & 1.028e$-$3 & 9.698 \\
\bottomrule
\multicolumn{5}{l}{*Bold font indicates the best value for each metric.}
\end{tabular}}
\end{table}

\FloatBarrier

\section{Conclusion}
\label{sec:conclusion}
This study introduced SSEMG-Net, a spectrogram-based denoising model for removing ECG artifact from sEMG. By integrating a physiologically aware sub-band masking strategy with a bidirectional Mamba backbone and dual-head decoding, SSEMG-Net effectively suppresses low-frequency ECG artifacts while preserving mid- and high-frequency sEMG content. Experimental results demonstrate that SSEMG-Net exhibits consistent improvements over existing waveform-based denoising approaches, achieving higher signal quality and more accurate feature preservation. These findings confirm that spectrogram-domain modeling, when combined with physiological priors, provides a powerful framework for sEMG enhancement. Future work will explore extending SSEMG-Net to address other types of sEMG contaminants and adapt it for streaming inference to enable real-time applications.

\bibliographystyle{IEEEtran}
\bibliography{refs}

\end{document}